\documentclass[aps,prb,amsfonts,amssymb,twocolumn,showpacs]{revtex4}
\usepackage{mathrsfs}
\usepackage{graphicx,hyperref}

\begin{document}
\title{All-optical switch with two periodically  curved nonlinear
 waveguides}
\author{Qiongtao Xie}
\affiliation{Institute of Physics, Chinese Academy of Sciences,
Beijing 100080, China}
\author{Xiaobing Luo}
\affiliation{Institute of Physics, Chinese Academy of Sciences,
Beijing 100080, China}
\author{Biao Wu}
 \affiliation{Institute of Physics, Chinese Academy of Sciences,
 Beijing 100080, China}
\email{bwu@aphy.iphy.ac.cn}
\begin{abstract}
We propose a type of all-optical switch which consists of two
periodically curved nonlinear optical waveguides placed in parallel.
 Compared to the all-optical switch based on the traditional nonlinear
directional coupler with straight waveguides, this all-optical
switch has much lower switching threshold power and sharper
switching width.
\end{abstract}
\maketitle

\section{Introduction}
The nonlinear directional coupler (NLDC), a device consisting of two
parallel straight nonlinear waveguides, has received much attention
for its potential applications as a type of all-optical switching
device\cite{Jensen,Maier,Agrawal}. Its switching operation is based on the
intensity-dependent power transfer between its two coupled
waveguides\cite{Gusovskii}. For a continuous-wave (CW) laser beam,
the beam can be directed towards different output ports of the NLDC
depending on whether the input power exceeds a threshold (or
critical) power $P_c$. At low input power, most of the light emerges
from the neighboring waveguide; at high power above $P_c$, most of
the light remains in the launching
 waveguide.
Therefore, the change in the input power can cause light to be
switched  from one waveguide to the other.

However, the application of the NLDC as an all-optical switch is
limited by its high threshold on switching power. There were two
approaches to improve its performance or to make it applicable. One
approach is to use a pulsed beam to lower the overall demand on the
laser power as a CW beam would. Both picosecond and femtosecond
lasers have been used to successfully demonstrate the all-optical
switching in a NLDC\cite{Friberg1,Friberg2,Weiner}. But experimental
results also showed some drawbacks. For example, an optical pulse
usually breaks up at the output ports and the switching is not as
sharp as the case of CW beams. The other is to use materials with
relatively larger Kerr nonlinearity but this usually leads to a
slower response time\cite{Vigil,Marchese}.

\begin{figure}[!tb]
\center
\includegraphics[width=0.5\textwidth]{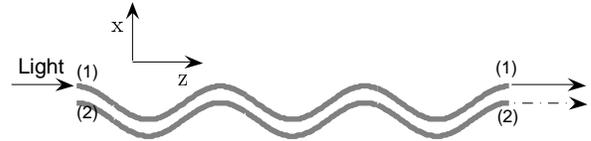}
\caption{Schematic drawing of a periodically curved all-optical
switch. The two curves represent two waveguides. The laser light
switches its output waveguide when the laser power is changed from
above to below a threshold value.} \label{fig1}
\end{figure}

In this paper, we propose a  different type of all-optical switch
with much lower threshold on switching power.  As illustrated in
Fig.\ref{fig1}, this device is made of two periodically curved
nonlinear waveguides placed
in parallel. It can be regarded as a modification of the traditional
NLDC which consists of two straight waveguides. Due to the effect of
nonlinear coherent destruction of tunneling (NCDT)\cite{Luo},
this modified NLDC can function as an all-optical switch at an
appropriate length.  This all-optical switch has lower critical power
since the periodical bending of the nonlinear waveguides can effectively
increase their nonlinearity. In addition, this device has sharper
switching width. Such a curved switch permits in principle arbitrarily
low switching threshold power; its only foreseeable drawback is longer
coupling length. This work is motivated by a recent experiment with  two
coupled periodically curved linear waveguides\cite{Della} and the
theoretical extension to the nonlinear case\cite{Luo}.

\section{Straight all-optical switch}
Before we fully discuss our proposed switch, we give a brief
 introduction to
the traditional NLDC. This device is made  of two straight nonlinear
 waveguides which
are placed in parallel adjacent to each  other so that they are
coupled
 optically.
The operation of the NLDC as an all-optical switching was studied in
detail by Jensen\cite{Jensen}. He showed that if all the
 light is
initially shined into one  waveguide  with the input power $P_0$,
then the amount of the light remaining in the launching waveguide is
given by \cite{Jensen,Maier}
\begin{equation}
P_1(L)=\frac{P_0}{2}[1+{\rm cn}(\pi L/L_c |m)]\,, \label{eq1}
\end{equation}
where $L$ is the length of the coupler and  ${\rm cn}(\pi L/ L_c|m)$
is the Jacobi elliptic function\cite{Jacobi}.  The parameter $L_c$ is
called the coupling length
representing the shortest length for the light switching from one
waveguide to the other without nonlinearity.  The other parameter
$m$ is defined as
$m=(P_0/P_c)^2$ with $P_c$ being the threshold (or critical) power
and given by
\cite{Friberg1,Friberg2,Weiner}
\begin{equation}
P_c=\lambda \sigma_{\rm eff}/L_cn_2,.
 \label{eq2}
\end{equation}
Here $\lambda$ is the free space wavelength of the light,
$\sigma_{\rm eff}$ is the effective cross-section of the waveguide,
and $n_2$ is the Kerr nonlinear coefficient (or nonlinear refractive
 index ).
The analysis of  the elliptic function cn shows that the NLDC can
function as a switch: most of light stays in the launching waveguide
if the input power $P_0$ is above the critical
power $P_c$ and  switch to the neighboring waveguide if $P_0<P_c$.

\section{Periodically curved all-optical switch}
Our proposed all-optical switch is a modification of the NLDC by
 replacing the two
straight waveguides with  two periodically curved nonlinear
waveguides as shown in Fig. \ref{fig1}. In the following discussion,
without loss of generality we focus
on the case where the two waveguides are bent  sinusoidally along
the propagation
direction $z$.  Specifically, the profile of the waveguide is given
by $x_0(z)=A\cos(2\pi z/\Lambda)$ with an amplitude $A$ and a period
$\Lambda$.

Since the light  is strongly localized in  the $y$
direction\cite{Della}.  In this case, the light propagation in this
nonlinear directional waveguides is described by an effective
two-dimensional wave equation \cite{Della,Luo,Micallef}
\begin{equation}
i\frac{\lambda}{2\pi}\frac{\partial \psi}{\partial
z}=-\frac{\lambda^2} {8\pi^2 n_{s}}\frac{\partial^{2} \psi}{\partial
x^2}+V[x-x_{0}(z)]\psi-n_2|\psi|^2\psi\,, \label{eq3}
\end{equation}
where  $V(x)\equiv [n_s^2-n^2(x)]/(2 n_s)\simeq n_s-n(x)$, where
$n(x)$ and $n_s$ are,
respectively, the effective refractive index profile of the
waveguides and the substrate refractive index. For the coupled
waveguides, $n(x)$ and thus $V(x)$ have a double-well structure. The
scalar electric field is related to $\psi$ through
$E(x,z,t)=(1/2)(n_s\epsilon_0 c_0/2)^{-1/2}[\psi(x,z)\exp(-i k c_0
t+ikn_sz)+c.c.]$, where $k=2\pi/\lambda$, $c_0$ is the speed of
light, and $\epsilon_0$ is the dielectric constant in vacuum. The
light intensity $I$ (in $W/m^2$) is given by
 $I=|\psi|^2=(n_s\epsilon_0 c_0/2)|E|^2$.
With the  Kramers-Henneberger transformation\cite{Henneberger},
$x'=x-x_0(z), z'=z$, and
$\phi(x',z')=\psi(x',z')\exp[-i(2n_s\pi/\lambda)\dot{x}_{0}(z')x'-
i(n_s\pi/\lambda)\int_{0}^{z'}d\xi\dot{x}_{0}^{2}(\xi)]$ (the dot
indicates the derivative with respect to $z'$), we have
\begin{equation}
i\frac{\lambda}{2\pi}\frac{\partial \phi}{\partial z'}=
H_0\phi-n_2|\phi|^2\phi+x'F(z')\phi, \label{eq4}
\end{equation}
where $H_0=-\frac{\lambda^2} {8\pi n_{s}}\frac{\partial^{2}
}{\partial x'^2}+V(x')$ and  $F(z')=n_s\ddot{x}_0(z')=(4\pi^2
 An_s/\Lambda^2)\cos(2\pi
z'/\Lambda)$ can be regarded as a ``force''  induced by the bending
 waveguide.

Due to the double-well structure of $V(z')$, we apply the two-mode
approximation\cite{Della, Luo} and write
 $\phi(x',z')=e^{-\frac{2i\pi}{\lambda}
E_0z'}[c_1(z')u_1(x')+c_2(z')u_2(x')]$, where $u_1$ and $u_2$ are
localized waves in the two waveguides and the two coefficients are
normalized to one, $|c_1|^2+|c_2|^2=1$. $E_0$ is defined as
$E_0=\int u_{1,2}^*H_0u_{1,2}dx'$. It is reasonable to assume that
the localized wave is a Gaussian, $u_{1,2}(x')=\sqrt{D}\exp[-(x'\pm
a/2)^2/2b^2]$, where $a$ is the distance between the two waveguides,
$b$ is the half-width of each waveguide, and $D$ is related to the
input power of the system $P_0$ as $D=P_0/(\sqrt{\pi}b)$. $P_0$ has
the unit of $W/m$. This two-mode approximation eventually simplifies
Eq.(\ref{eq4}) to
\begin{eqnarray}
i\dot{c_1}&=&\frac{v}{2}c_2-\frac{S}{2}\cos(\omega
z')c_1-\chi|c_1|^{2} c_1,
\label{eq5}\\
i\dot{c_2}&=&\frac{v}{2}c_1+\frac{S}{2}\cos(\omega
z')c_2-\chi|c_2|^{2} c_2\label{eq6},
\end{eqnarray}
where $S=8\pi^3aAn_s/\Lambda^2\lambda$, $v=4\pi(\int
u_1^*H_0u_2dx)/\lambda$,  $\omega=2\pi/\Lambda$, and
$\chi=\sqrt{2\pi}
 n_2P_0/(\lambda b)$.
Since the real waveguide is 3D, we replace $b$ in $\chi$ with
 $\sigma_{\rm eff}$
to relate our nonlinear parameter $\chi$ to real experimental
 parameters and write
$\chi=2\pi n_2P_0/\lambda\sigma_{\rm eff}$ \cite{Jensen, Maier},
where $P_0$ has the unit of $W$.  When $S=0$,
Eqs.~(\ref{eq5}\&\ref{eq6}) are reduced to the well-known Jensen
equation, where the critical power is defined as $P_c=2vP_0/\chi$
and the coupling length $L_c$ is related to $v$ through
$L_c=\pi/v$\cite{Jensen}.  Combination of these relations leads to
the critical power $P_c$ in Eq.(\ref{eq2}).
\begin{figure}[htb]
\begin{center}
\includegraphics[width=0.5\textwidth]{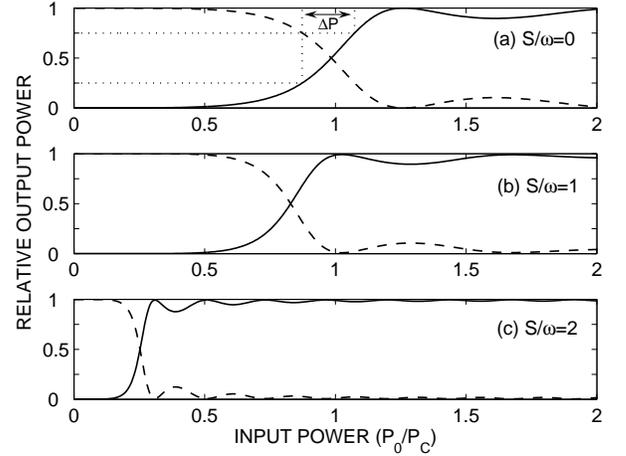}
\caption{Relative output powers as functions of the input power for
three different values of $S/\omega$ at $\omega/v=10$. The solid
lines are for the launching waveguide while the dashed lines are for
 the
neighboring waveguide. The input power $P_0$ is scaled by the
threshold switching power $P_c$. } \label{fig2}
\end{center}
\end{figure}

\begin{figure}[tb]
\begin{center}
\includegraphics[width=0.5\textwidth]{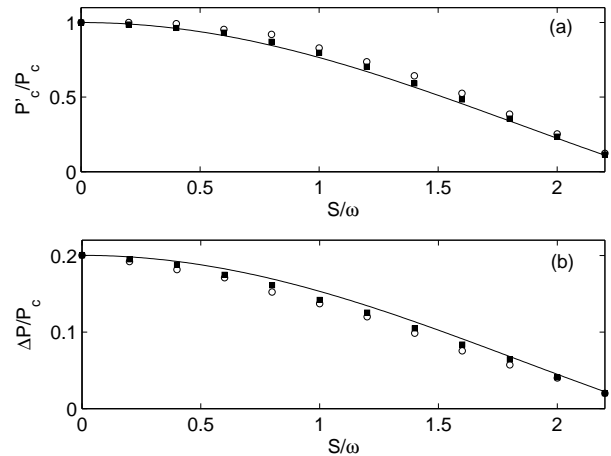}
\caption{(a)Threshold switching power as a function of the ratio
 $S/\omega$.
(b) Switching width as a function of the ratio $S/\omega$.  The open
circles are for $\omega/v=10$, the squares are for $\omega/v=20$,
and the solid lines represent the theoretical results at the high
frequency limit.} \label{fig3}
\end{center}
\end{figure}

The presence of the periodical curvature in the two coupled
waveguides strongly affect the behavior of the all-optical switch.
To investigate this effect, we solve  Eqs. (\ref{eq5}) and
(\ref{eq6}) numerically. Figure \ref{fig2} shows the relative output
power (to the total power) as a function of the input
 power for three
different values of the ratio $S/\omega$ with $\omega/v=10$. We
observe two trends as the ratio $S/\omega$ is increased. First, the
threshold switching  power $P'_c$ decreases. Second, the width of
 the
switching step becomes smaller.  This demonstrates that by
increasing the amplitude of the periodic curvature one can improve
the performance of the all-optical switch in two aspects: lowering
the threshold switching power and sharpening the switching. We have
computed numerically how the critical switching power changes with
$S/\omega$ for two different values of ratio $\omega/v$. The results
are plotted in Fig.\ref{fig3}(a), showing a significant lowering of
threshold switching power as $S/\omega$ increases. To measure the
width of the switching steps in
 Fig.\ref{fig2},
we introduce a new quantity $\Delta P$, which is the distance
between
 the
two positions of the input power where the relative output powers
are $25\%$ and $75\%$, respectively. Figure \ref{fig3}(b) shows that
the switching width $\Delta P$ becomes smaller with increasing
 $S/\omega$,
i.e., the switching becomes sharper and sharper as $S/\omega$
 increases.

To better understand the above results, we consider the high
frequency limit, $\omega \gg \max\{v,\chi\}$, which is usually the
case for
 current
experiments with optical waveguides. We take advantage of the
transformation $c_{1,2}=c^\prime_{1,2}\exp[\pm iS\sin(\omega
z')/2\omega]$. After averaging out the high frequency terms,
we arrive at a non-driving nonlinear model \cite{Luo,Longhi},
\begin{eqnarray}
i\dot{c_1'}&=&\frac{v}{2}J_0(S/\omega)c_2'-\chi|c_1'|^{2} c_1',
\label{highfrequency1}\\
i\dot{c_2'}&=&\frac{v}{2}J_0(S/\omega)c_1'-\chi|c_2'|^{2}
c_2'\label{highfrequency2},
\end{eqnarray}
where $J_0(x)$ is the zeroth-order Bessel function\cite{Jacobi}.
These
 two
equations are exactly the Jensen equations\cite{Jensen} except that
the coupling constant $v$ is renormalized by a factor of
$J_0(S/\omega)$. This shows that all the effect of the periodic
bending is manifested in the renormalized factor $J_0(S/\omega)$.
Consequently, the nonlinear parameter $\chi$ is effectively
increased by a factor of $1/J_0(S/\omega)$ and the threshold power
$P'_c$ is lowered by a factor of $J_0(S/\omega)$
\begin{equation}
P'_c=\lambda \sigma_{\rm eff}J_0(S/w)/L_cn_2=P_cJ_0(S/\omega)\,.
\label{eq10}
\end{equation}
This implies that arbitrarily low switching powers may be obtained
when the ratio $S/\omega$ is chosen to be close to the first zero of
the Bessel function $J_0(S/\omega)$. This analytical result is
compared to our previous numerical results in Figure \ref{fig3}(a)
and they match very well. The switching width can also be computed
by combining Eq. (\ref{eq10}) and Eq.(\ref{eq2}). The results agree
very well  with our previous numerical results as seen in Fig.\ref{fig3}(b).

As in the traditional NLDC, an appropriate length has to be chosen
for our device to function as an all-optical device. For the
traditional NLDC, the length is the coupling length $L_c$. For our
device, the appropriate length, which may be called switch length,
is given by
\begin{equation}
L'_c=L_c/J_0(S/\omega)\,. \label{eq11}
\end{equation}
Together with Eq. (\ref{eq10}), this indicates that to lower the
 threshold
switching power we have to make a sacrifice by using longer
waveguides. In fact, if one wants to lower the critical switching
power by a
 factor, then
one has to use waveguides which is longer by the same factor.

As an example, we take the experimental parameters in Ref.
\cite{Friberg2}  to estimate our theoretical values in  Eqs.
(\ref{eq10}) and (\ref{eq11}) for a given ratio of $S/\omega$. In
this experiment, the wavelength of the laser light is
 $\lambda=0.62\mu$m,
the effective cross-section area of the waveguide is $\sigma_{\rm
eff}=15\mu$m$^2$, the nonlinear index $n_2=3.2\times 10^{-16}{\rm
cm}^2/{\rm W}$,  and the coupling length  $L=L_c= 5$mm. The critical
power is thus $P_c=60$ kW. For the periodically bent coupler, we
choose $S/\omega \approx 2.38$, which means $J_0(2.38)\approx 0.01$.
As a result, the threshold switching power is lower by a factor of
100 and becomes $P'_c=600$W while the switch length of the coupler
is $L'=0.5$m.

\section{Conclusion}
In conclusion, we have proposed a modification to the traditional
NLDC by bending the nonlinear waveguides periodically. When this
device functions as an all-optical switch, it has much lower
threshold switching power and sharping switching.

\section*{Acknowledgments}
This work is supported by the ``BaiRen'' program of Chinese Academy
of Sciences, the NSF of China (10504040), and the 973 project of
China(2005CB724500,2006CB921400).

\end{document}